\documentclass[a4paper]{article}
\pdfoutput=1
\usepackage[latin1]{inputenc}
\usepackage{xspace}
\usepackage[pdftex]{graphicx}
\usepackage{url}
\usepackage{subfig}

\usepackage{listings}
\usepackage{xcolor}
\definecolor{Gray}{gray}{0.70}
\usepackage[
	pdfauthor={Jan Wolter},
	pdftitle={Visual Representation of 3D Language Constructs Specified by Generic Depictions},
	pdfsubject={Technical Report},
	pdfkeywords={},
	pdfcreator={pdflatex},
	colorlinks=false,
	linkcolor={black},
	citebordercolor={white},
	linkbordercolor={white},
	urlbordercolor={white},
        bookmarksopen={true}
]{hyperref}

\usepackage{nccfoots}
\usepackage{url}

\newcommand{\dev}{DEViL3D\xspace}
\newcommand{\dotcup}{\mathbin{\dot{\cup}}}

\begin{document}

\begin{center}
\textbf{\Large  Visual Representation of 3D Language \\[3pt]Constructs Specified by Generic Depictions}\\[10pt]
  Jan Wolter \\
  University of Paderborn \\
  Department of Computer Science  \\
  F\"{u}rstenallee 11, 33102 Paderborn, Germany   \\
  jan.wolter@uni-paderborn.de
\end{center}

\begin{abstract}

Several modeling domains make use of three-dimensional representations, e.g., the ``ball-and-stick'' models of molecules. Our generator framework \dev supports the design and implementation of visual 3D languages for such modeling purposes. The front-end of a language implementation generated by \dev is a dedicated 3D graphical structure editor, which is used to construct programs in that domain. \dev supports the language designer to describe the visual appearance of the constructs of the particular language in terms of generic 3D depictions. Their parameters specify where substructures are embedded, and how the graphic adapts to space requirements of nested constructs. The 3D editor used for such specifications is generated by \dev, too. In this paper, we briefly introduce the research field of 3D visual languages and report about our generator framework and the role that generic depictions play in the specification process for 3D languages. Our results show that our approach is suitable for a wide range of 3D languages. We emphasize this suitability by presenting requirements on the visual appearance for different languages.

\Footnotetext{}{This paper is a extended version of my paper \emph{Specifying Generic Depictions of Language Constructs for 3D Visual Languages}, which I presented at the VL/HCC 2013: \url{http://dx.doi.org/10.1109/VLHCC.2013.6645258}}

\end{abstract}

\subparagraph{Key words.} three-dimensional depictions, visual languages, visual programming, automated generation, 3D interaction techniques.

\tableofcontents
\pagebreak

\section{Introduction}
\label{sec:intro}
Visual languages are particularly beneficial for domain-specific applications, since they support graphical metaphors of their domain. Up to now the majority of visual languages are two-dimensional. Examples are LabVIEW \cite{labview}---used in industrial automation and instrument control---and the well-known UML, which is used to model several aspects of object-oriented software systems. Both languages use two-dimensional representations, e.g., boxes and lines connecting them, in order to visualize dataflow or dependences.

For some domains, using the third dimension is advantageous or necessary: Inherently three-dimensional languages that make use of real-world objects as architecture-like modeling domains can be represented without loss of information only in 3D. The ``ball-and-stick'' models of molecules visualize atoms as balls and bonds between them as sticks. The arrangement of the atoms in 3D space is the result of the electron cloud repulsion and therefore the arrangement of language constructs to one another is inherently 3D.

Another argument in favor of 3D languages is to assign a semantic meaning to each dimension. A good example is the web-based 3D editor \emph{ToneCraft} \cite{tonecraft} that lets the user build music in 3D by composing boxes into a matrix-like area: the $\mathrm{y}$-axis represents the pitch of a tone, the $\mathrm{x}$-axis represents the time and the $\mathrm{z}$-axis makes it possible to layer sounds. Even \emph{Petri Nets} can benefit from a three-dimensional representation, for example, Petri Nets modeling different aspects---such as control flow and data flow---that have connections to one another. The representation of such Petri Nets in 2D is confusingly complex due to crossing edges and confound aspects. This can be solved by laying out the Petri Net in three dimensions, where each aspect is represented on a different plane, which can be stacked together to one Petri Net \cite{Roel07}.

Moreover, the third dimension can be used to overcome limitations in 2D arrangements. For example, in some cases the 2D representation of UML diagrams is not efficient enough and can be extended to 3D, e.g., to overcome the problem of intersecting edges in sequence diagrams. Alternatively, the third dimension can be used to focus on specific classes of interest in class diagrams \cite{GRR99}.

Some further systems in context of visual programming make use of three-dimensionality.
\emph{Alice} \cite{KP07} is a 3D programming environment that employs a story telling metaphor and teaches children to learn fundamental programming concepts by programming the behavior of 3D objects, e.g., people or animals. This system's effectiveness has been proven by various user studies. Another programming environment that uses three-dimensional objects is \emph{AgentCubes} \cite{IRW09}, the successor of \emph{AgentSheets}, which allows children to create interactive 3D games. The key challenges of AgentCubes are intuitive mechanisms to create 3D objects incrementally, including subsequent programming and animation aspects.

The above mentioned 3D languages and programming environments consist of objects with different 3D shapes. Some languages---as molecular models, ToneCraft, or Petri Nets---use relatively simple shapes like cubes, spheres, or cylinders. But the 3D scenes composed with Alice or AgentCubes consist of more complex shapes, visualizing real world objects such as people, buildings, or cars. There is an additional challenging task: Because of the fact that language constructs can be nested inside each other, their depiction needs mechanisms to adopt the size of interior constructs.

Our approach relies on a tool that makes the development of 3D language editors as simple as possible. The development of a language specific implementation is justified, only if the effort is appropriately small. Therefore, effective generator systems are useful. We are developing the generator framework \dev (\underline{D}evelopment \underline{E}nvironment for \underline{Vi}sual \underline{L}anguages in \underline{3D}) that accomplishes this task. One central part of developing visual languages (either two- or three-dimensional) is the definition of the visual appearance of language constructs. For such a task, \dev provides a 3D editor to specify \emph{generic depictions} for language constructs. They may consist of a large set of three-dimensional geometric shapes and their parameters specify where substructures are embedded, and how the graphic adapts to space requirements of nested constructs.

\begin{figure}[!ht]
  \centering
  \includegraphics[width=10cm]{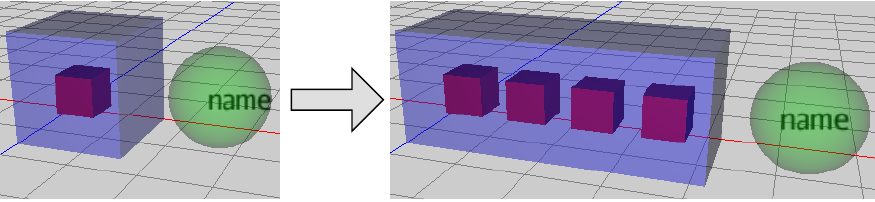}
  \caption{A language construct before and after stretch.}
  \label{fig:depicStretch}
\end{figure}

Figure \ref{fig:depicStretch} shows an exemplary language construct consisting of a blue box and a green sphere. The green sphere contains a text label and the blue box is able to embed further language constructs visualized as red boxes. These language constructs are laid out as a list, which grows along the $\mathrm{x}$-axis. When new constructs are inserted, the blue box has to adapt its space requirements to the needs of the nested list and the green sphere has to move right.

The remainder of this paper is structured as follows. In Section \ref{sec:devil3d}, we give a brief overview of \dev. Since this paper focuses on generic depictions of 3D languages, Section \ref{sec:genDepic} introduces these and presents an interactive 3D language editor to compose them and illustrates a method instantiated depictions perform to adapt their size according to embedded constructs. In Section \ref{sec:rangeOfAppl} we show that specifying generic depictions according to our approach is applicable for a large set of languages. We will give a survey over developing generic depictions for different languages. Then we discuss related work and compare it to our approach. Section~\ref{sec:conclusion} concludes the paper.

\section{\dev}
\label{sec:devil3d}
The generator system \dev \cite{Wol12} allows to generate 3D structure editors, supporting the \emph{direct manipulation} paradigm~\cite{Shn83} and, therefore, prevent the user from constructing syntactically incorrect programs. \dev combines approved concepts of the predecessor system DEViL \cite{SKC06, SCK07, SK03} and new aspects necessary to construct three-dimensional programs. This section gives a brief overview on the steps elementary to generate a structure editor with DEViL3D. It includes an overview on several specification parts and, additionally, a brief outline on 3D-specific aspects that are applicable in all generated editors.

\begin{figure}[!ht]
  \centering
  \includegraphics{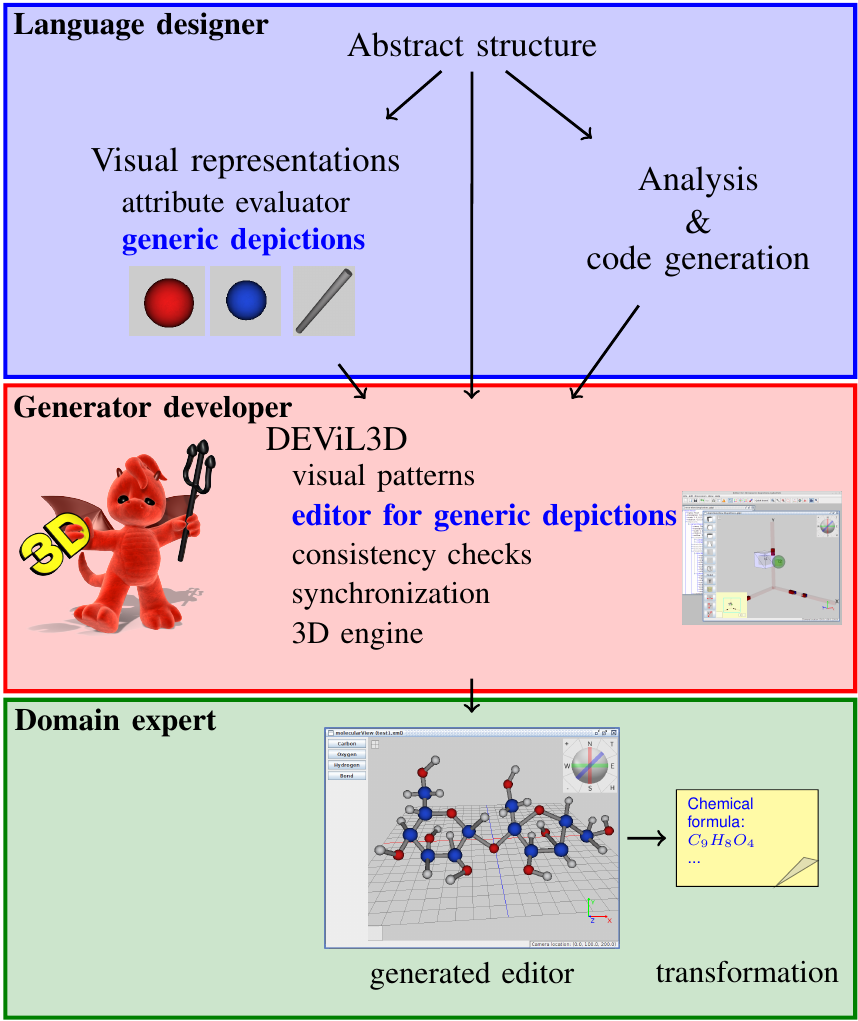}
  \caption{Specification process with \dev.}
  \label{fig:devil3dSpecification}
\end{figure}

Figure~\ref{fig:devil3dSpecification} visualizes the specification process. It is layered into three parts. The upper area shows specification parts a language designer has to formulate: the \emph{abstract structure}, \emph{visual representations}, and \emph{code generators}. The area in the middle illustrates our generator framework \dev. As input, \dev gets a language specification and generates a language processor which has a dedicated 3D graphical structure editor as its front-end. Domain experts use such editors to construct three-dimensional programs of their domain, e.g., molecular models.

The abstract structure describes the language constructs and how they are connected, without defining a concrete representation. For such purpose, a specifically tailored textual domain specific language is used, which is strongly related to object oriented programming languages. It is based on well known concepts like classes, inheritance, attributes, and references.

The specification of the visual representation is based on attribute grammars, which are translated into computations of the graphical representation and arrange objects in 3D space. The attribute grammars are based on a context-free grammar that is generated from the abstract structure. The arrows in Figure \ref{fig:devil3dSpecification} indicate this dependency. \dev provides a library of \emph{visual patterns} that encapsulate common representation arrangements like three-dimensional sets, lists, line connections, or cone-trees. They are defined as visual roles, which can be assigned to symbols of the grammar in a declarative way. The visual patterns automatically contribute layout and interaction properties. The generic depictions, which define the visual representation of a language construct, are referenced by visual patterns. For the generation of a molecular editor with \dev, amongst others, generic depictions for kinds of atoms and for bonds are needed. They play the role of building blocks from which the visual program will be composed.

Further to the visual representation, the language designer can define a set of code generators based on attribute grammars that transform the 3D program into different textual representations.

To ensure that an editor generated by \dev is as easy as possible to use, we have analyzed established three-dimensional editors from different domains, e.g., Autodesks \emph{3ds Max} \cite{3dsmax}, which allows to create 3D models and animations. From this and other systems, we have adopted basic techniques to make the interaction with the 3D scene as simple as possible. Such challenge will become clear, if we compare editors that allow the creation of 2D and 3D representations. They can be distinguished by the \emph{degree of freedom} (DOF) that describes the possibilities of object placement in space. The 2D space has three DOFs: translation along the two axes and rotation around the neutral point. In contrast, there are twice as many DOFs in three-dimensional space, namely, translation and rotation along all three axes.

In order to cope with such an increase in complexity, editors generated by \dev comprise techniques for navigation, interaction, and layout. Our aim is that the editors are usable with a classical mouse, which can not directly capture all six DOFs. To support this, dedicated \emph{widgets} are provided that perform interaction tasks like translating, scaling, or rotating. In complex 3D scenes objects can occlude each other, but a first-person-view camera lets the user navigate inside the scene and explore it.

The layout and interaction tasks are encapsulated in visual patterns. The language designer does not need to implement such functionality. The assignment of visual patterns to language constructs is generally sufficient. Interaction tasks are automatically tailored to the needs of the representation of a visual pattern: For example, elements of a linearly ordered list can only be moved along the direction of its arrangement. The task of inserting language constructs is triggered by so called \emph{insertion contexts} that are also tailored to the needs of the visual pattern.

The implementation of 3D-specific aspects makes use of the underlying \emph{jMonkeyEngine} \cite{jme}, which in turn uses \emph{OpenGL} to render 3D scenes. To organize the 3D scene efficiently, the provided \emph{scene graph}---a data structure to organize the 3D scene---is used extensively.

\section{Generic depictions}
\label{sec:genDepic}
Generic depictions describe the visual appearance of constructs of a particular language in terms of generic 3D graphics.
From a formal point of view generic depictions are an abstract concept that can be described by a quadruple of graphical primitives, representation properties, containers, and stretch intervals: $\mathcal{D} = (\mathcal{P}, \mathcal{R}, \mathcal{C}, \mathcal{I})$. These parts carry out the following tasks:

\begin{itemize}
  \item A collection of graphical primitives $\mathcal{P}$ determine the graphical representation of a language construct. The following types of primitives are available:  $\mathcal{P}$ determine the graphical representation of a language construct:  $\mathcal{P} = Box \dotcup Sphere \dotcup Cone \dotcup Cylinder \dotcup Arrow \dotcup Line \dotcup Quad \dotcup Torus \dotcup 3DModel \dotcup Text$. 3D models allow to integrate objects with more complex shapes.
  \item The representation properties $\mathcal{R}$ describe materials, e.g., color or texture definitions, that can be mapped to graphical primitives~$\mathcal{P}$.
  \item A set of containers $\mathcal{C}$ is responsible to embed nested objects of arbitrary size, when the generic depiction is instantiated. Each container needs a unique name.
  \item The specification of layout properties is managed by a set of stretch intervals $\mathcal{I}$. Such intervals determine, which part of a container grows, when the size of nested objects exceeds the container's size.
\end{itemize}

\begin{figure}[!ht]
  \centering
  \includegraphics[width=\columnwidth]{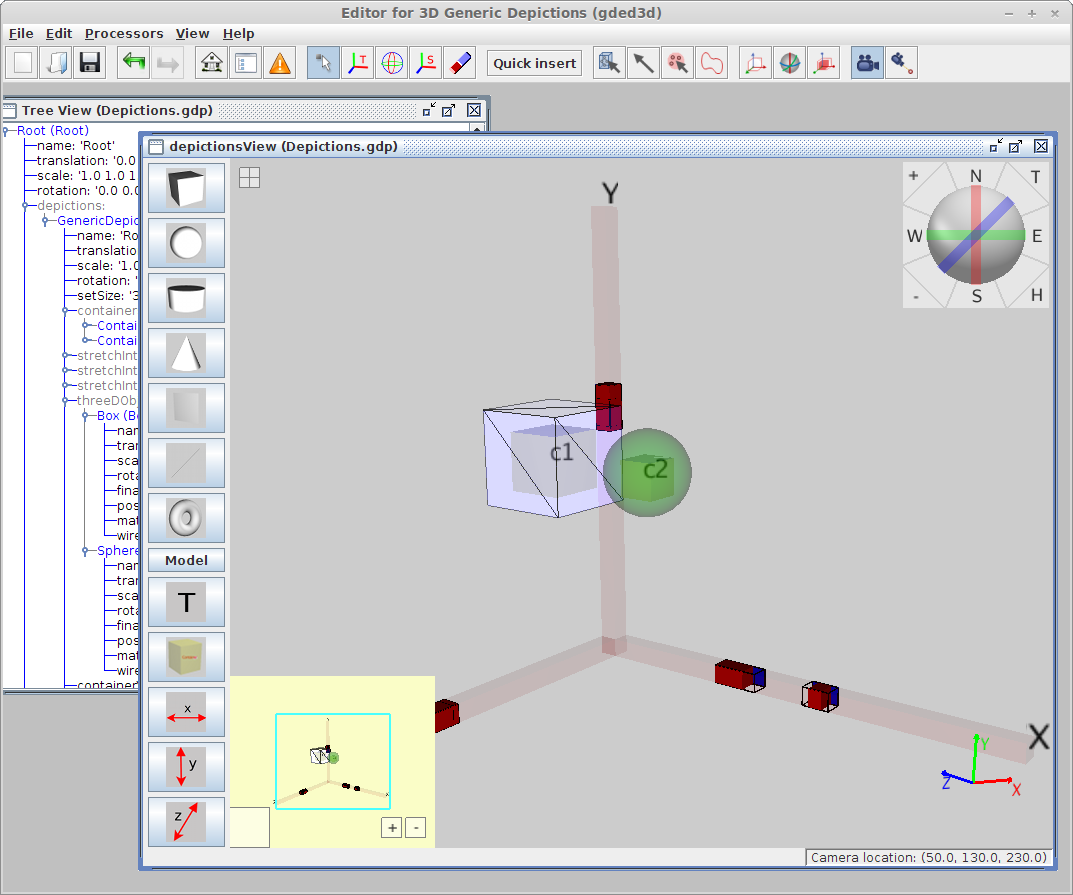}
  \caption{3D editor for generic depictions.}
  \label{fig:gded3d}
\end{figure}

The best way to build visual constructs is to do it visually. Such an approach ensures a close mapping between the domain world and the notation (according to the cognitive dimension \emph{closeness of mapping} \cite{GP96}). Figure \ref{fig:gded3d} shows a screenshot of the generic depiction editor that provides a 3D canvas in which the language designer visually composes a set of generic depictions for some language constructs. The depiction shown in the figure consists of two graphical primitives in form of a box and a sphere, two containers named $c1$ and $c2$, and four stretch intervals. One container is located inside the box, the other one in the sphere. Each stretch interval is responsible for one dimension and is located on light red colored coordinate axes. The abstract visual program in Figure \ref{fig:depicStretch} uses the generic depiction specified in Figure~\ref{fig:gded3d}.

Containers constitute the interface of generic depictions. If a language designer wants to change the visual representation of a language construct, two generic depictions can replace each other, if they coincide according to the number and the naming of their containers.

The following requirements restrict a semantically correct generic depiction: Each container must be covered by at least one stretch interval in each dimension. Otherwise, it is not clear how to respond to increasing size requirements of nested constructs. Any two stretch intervals must not overlap, so that in all cases a uniquely determined interval is responsible to stretch the container.

The following subsection will convey an impression, how the stretch algorithm for instantiated depictions proceeds. Part \ref{subsec:genDepicEditor} presents how generic depictions can be constructed with our 3D editor. Afterwards, we explain the code generation process that transforms the 3D program into Java code. We have developed the 3D editor for generic depictions by pursuing a \emph{bootstrapping approach}. The editor is specified with \dev. Hence, we have acted as a language designer and developed a set of high-level specifications as presented in Figure~\ref{fig:devil3dSpecification}, which describes the 3D language for generic depictions. Of course, visual representations of language constructs used in this editor have initially been implemented on a lower level, i.e. textual descriptions and Java code. This specification part is comparable with the code the editor for generic depictions generates within \dev.

\subsection{Application of Generic Depictions}
As seen in Figure \ref{fig:depicStretch} language constructs of 3D visual programs---which support embedded substructures---have to adopt their size according to the requirements of these substructures. Such nested structures require the specification of containers and stretch intervals as seen in Figure \ref{fig:gded3d}. In general, the embedded constructs can be laid out according to any visual pattern. In the example of Figure \ref{fig:depicStretch} the list pattern is used for the constructs inside the blue box.

By inserting more constructs in the list, the container's size could become insufficient. To adapt the size, an algorithm automatically stretches the container linearly. The algorithm operates on one depiction and iterates over all containers for each spatial dimension. If the preferred size, determined by the nested constructs, exceeds the actual container size, all stretch intervals that intersect the container are computed. The intervals of a container that are covered by a stretch interval will be stretched linearly. For this purpose, each container must be covered by at least one stretch interval.

\begin{figure}[!ht]
  \centering
  \includegraphics[width=11cm]{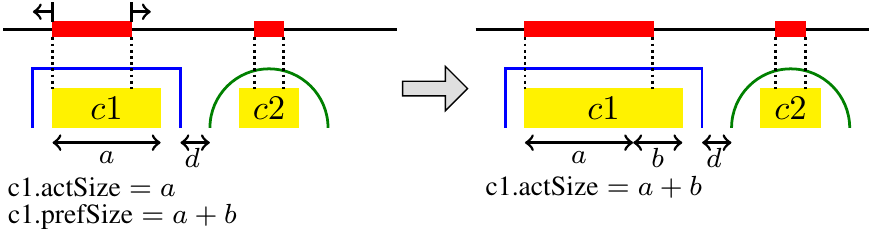}
  \caption{Behavior of the stretch algorithm.}
  \label{fig:stretchAlgoPic}
\end{figure}

Figure \ref{fig:stretchAlgoPic} shows a schematic sketch of the generic depiction of Figure \ref{fig:gded3d} reduced to the $\mathrm{x}$-axis to illustrate the stretch algorithm for one dimension. Initially the actual size of container $c1$ is $a$, but the embedded constructs need more space, so the preferred size is $a+b$. Hence, container $c1$ must be stretched to reach the new size. The stretch process behaves as if the containers and the primitives were printed onto elastic rubber and the start and end position of the stretch intervals were handles. To stretch the container and the primitives, the algorithm ``pulls'' the margins of the interval. Areas that are not covered by stretch intervals are not be stretched. In particular, the distance of two such points must remain the same after the transformation. Hence, the distance $d$ between the box and the sphere is the same after application of the stretch algorithm and the sphere with container $c2$ inside has simply to move right.

This algorithm is encapsulated in \dev and always comes into operation when new elements are inserted into a container. From an editor user's point of view, the algorithm provides a natural behavior when new language elements are inserted. Language designers do not need to care about the dynamic adaptation of generic depictions, as the generator system automatically provides it.

\subsection{The 3D Editor for Generic Depictions}
\label{subsec:genDepicEditor}
\begin{figure}[!ht]
 \begin{center}
  \subfloat[Cube-shaped insertion context to insert containers and primitives. \label{subfig:insCont3d}]{\includegraphics[width=10cm]{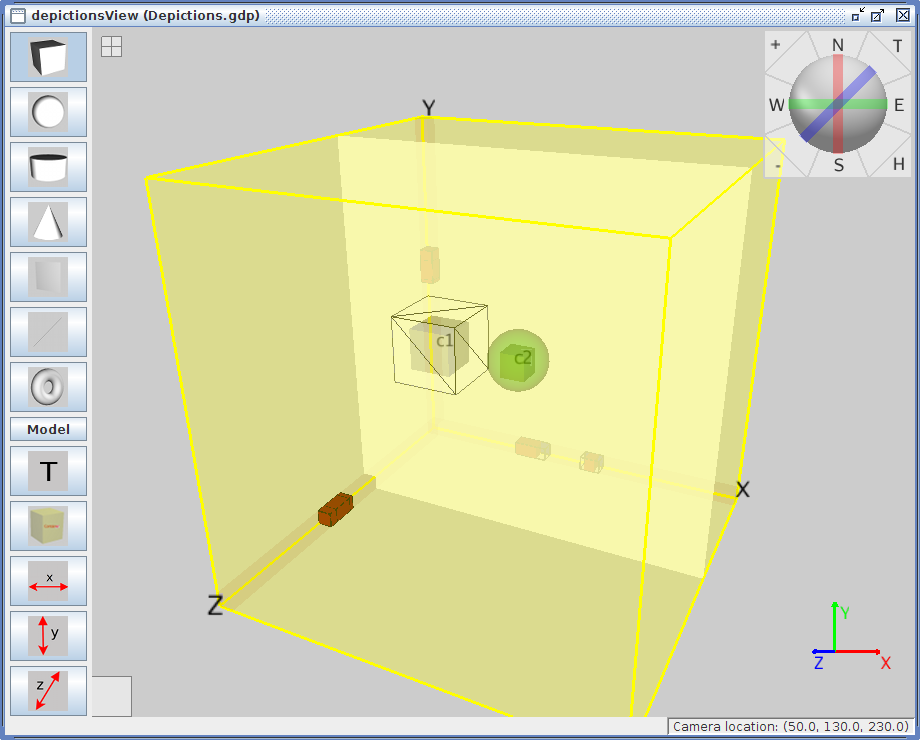}}\\
  \subfloat[Plane-shaped red insertion context to insert stretch intervals. \label{subfig:insCont1d}]{\includegraphics[width=10cm]{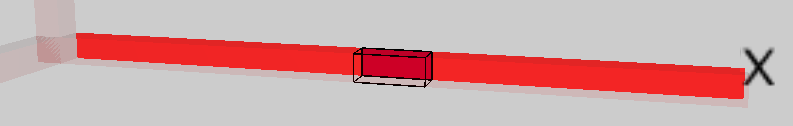}}
 \end{center}
\caption{Different kinds of insertion contexts.}
\label{fig:insCont}
\end{figure}

The editor for generic depictions provides a multi-document interface that shows the three-dimensional view, where a depiction is created (see Figure \ref{fig:insCont}\subref{subfig:insCont3d}). In the middle of the canvas the generic depiction can be constructed. The buttons on the left-hand side represents components of generic depictions (see item list at the beginning of Section \ref{sec:genDepic}) that can be inserted into the 3D view. When a button is clicked, all appropriate three-dimensional insertion contexts appear in the 3D canvas and highlight valid positions, where such an object can be inserted. The nature of an insertion context is determined by the way the language object is organized in the scene.

Since graphical primitives and containers can be placed anywhere in 3D space, the insertion context is cube-shaped and includes a plane to indicate a three-dimensional position (see Figure \ref{fig:insCont}\subref{subfig:insCont3d}). This plane can be moved along the $\mathrm{z}$-axis. The 3D position of the object is then determined by a mouse click on the plane. The stretch intervals can be positioned onto coordinate axes, which are located in the 3D scene. To insert such intervals, a plane-shaped insertion context appears along the chosen coordinate axis (see Figure \ref{fig:insCont}\subref{subfig:insCont1d}), whereby the user can determine a position on the axis.

\begin{figure}[!ht]
 \centerline{
  \subfloat[Translate 3D \label{subfig:translate3d}]{\includegraphics[height=3.2cm]{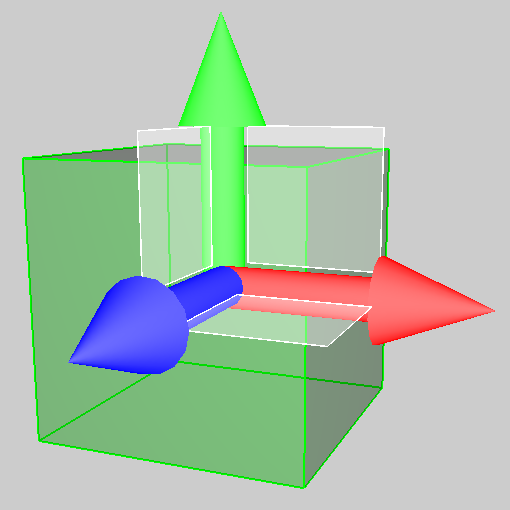}}
  \hfil
  \subfloat[Translate 1D \label{subfig:translate1d}]{\includegraphics[height=3.2cm]{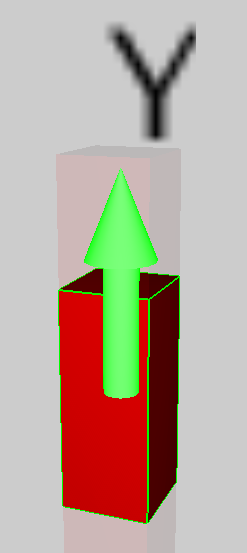}}
  \hfil
  \subfloat[Scale \label{subfig:scale}]{\includegraphics[height=3.2cm]{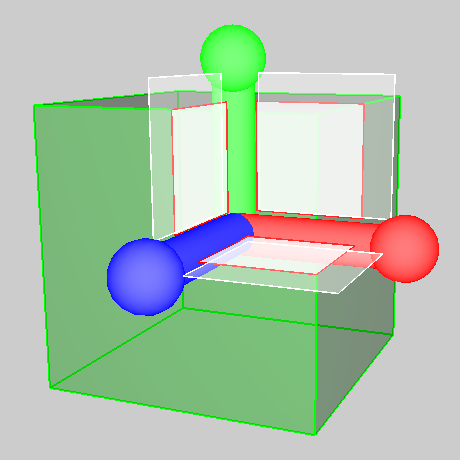}}
  \hfil
  \subfloat[Rotate \label{subfig:rotate}]{\includegraphics[height=3.2cm]{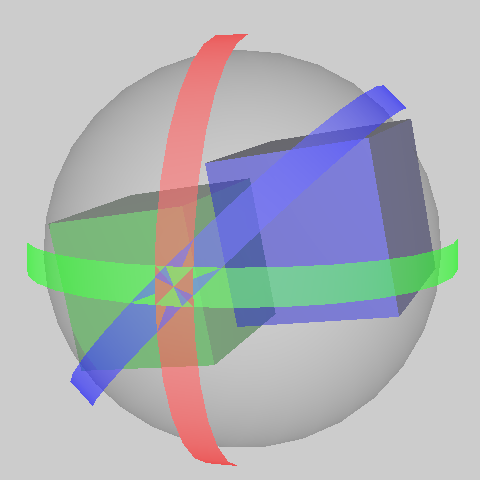}}
 }
\caption{Widgets to manipulate objects.}
\label{fig:widgets}
\end{figure}

After the insertion of a generic depiction's component, the language designer is able to modify it. In order to translate, scale, or rotate the object, a suitable widget attached to the object in question (see Figure~\ref{fig:widgets}). Such widgets are automatically adjusted to the requirements of the object according to the degree of freedom the object has. The graphical primitives and containers can be positioned everywhere inside the yellow insertion context and can therefore be manipulated along all three spatial dimensions. The widget to translate an object provides three arrows and three planes between them. The user may translate it along one dimension separately (via an arrow) or along two dimensions simultaneously (via a plane). However, stretch intervals can only be reshaped along one spatial dimension, as depicted in Figure \ref{fig:widgets}\subref{subfig:translate1d}.

These manipulation techniques are generally available in all structure editors generated with \dev. From a language designers point of view, the automatic adaptation of widgets is achieved by the application of visual patterns. Since the editor for generic depiction is bootstrapped, graphical primitives and containers are laid out according to the 3D set pattern and stretch intervals according to the 1D set pattern. The latter determines the position along two dimensions and lets the user only change the position along one continuous dimension.

\begin{figure}[!ht]
 \begin{center}
  \subfloat[Cylinder metaphor \label{subfig:cylinderSel}]{\includegraphics[height=6cm]{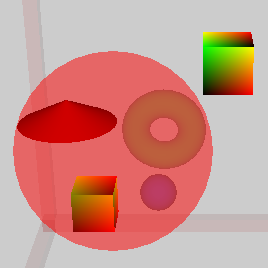}}
  \hfil
  \subfloat[Lasso metaphor \label{subfig:lassoSel}]{\includegraphics[height=6cm]{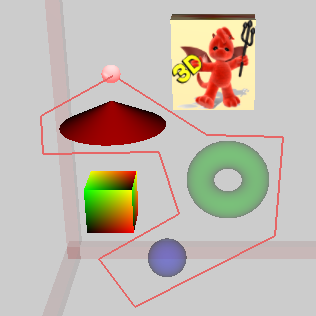}}
 \end{center}
\caption{Two ways to select multiple objects.}
\label{fig:multSelection}
\end{figure}

All interaction and manipulation tasks can be performed by using a classical 2D mouse. To do so, the editor supports a \emph{ray casting} metaphor. If the user wants to interact with an object, a ray starting at the mouse cursor is shot into the 3D scene to determine the first object that is intersected by the ray. This technique supports the direct manipulation approach in 3D and is used for all mentioned tasks: the interaction with insertion contexts to insert objects, the selection, and manipulation of language constructs with widgets. In some cases it is desirable to select multiple objects at the same time. The structure editor provides different methods to accomplish this task. As in many editors---2D or 3D---the user may select multiple objects one after the other by picking them while a special control key is pressed. There are 3D scenes where this approach is not applicable: Objects that are placed far away from the actual camera position appear much smaller than near located objects, which makes a precise selection difficult. To overcome this problem, we have developed a \emph{cylinder metaphor} inspired by Tavanti et al. \cite{TDL04}. The user of the editor sees only the circle-shaped cylinder cover on the 2D monitor screen (see Figure \ref{fig:multSelection}\subref{subfig:cylinderSel}). The cylinder expands into the 3D scene and selects all objects that are enclosed by it. Of course, the editor user can adjust the size of the lateral surface. But again, even this method is not distinct enough, if the objects intended for selection cannot be captured by a cylinder. A \emph{lasso metaphor} may be better suited: An arbitrary polygon can be created, which expands into the 3D scene as well to select the objects (see Figure \ref{fig:multSelection}\subref{subfig:lassoSel}).

Particularly important for the visual representation of generic depictions is, apart of the shape of the objects, their color. In the context of 3D applications the term \emph{material} has been established, which includes everything that influences what the surface of a 3D object looks like: the color, texture, shininess, and transparency. The generic depictions allow to define three different types of materials. Color materials that define RGB and transparency values. A texture material that refers to a normal image file to realize a special impression (e.g. the box in the top-right corner in Figure \ref{fig:multSelection}\subref{subfig:lassoSel} that refers to an image file of the \dev logo). And a custom material that refers to a shader specification, which defines sets of instructions that are executed on the GPU to realize special effects (see the box on the right-hand side in Figure \ref{fig:multSelection}\subref{subfig:cylinderSel}). The language designer and user of the generic depictions editor may create arbitrary many material objects, which are referred by almost all graphical primitives. The only exception are 3D models, which model very complex shapes and were created by external 3D modeling frameworks like Blender~\cite{Blender}.

Generally, 3D scenes can be more complex than 2D scenes and therefore objects may be hard to locate when they are far away. To overcome such problems the user needs mechanism to explore the 3D scene. To do so, a camera is used that allows to \emph{explore} the scene without an explicit target, or to \emph{search} a particular target. The editor offers different ways to control the camera. The \emph{ViewSphere} located on the upper right corner allows to rotate the camera around any language object to be selected by the user (see Figure \ref{fig:insCont}\subref{subfig:insCont3d}). Furthermore, the ViewSphere provides around its sphere different \emph{segments} labeled with a cardinal point to switch the viewing position rapidly and show the scene from eight distinct positions. To indicate the actual position, the editor provides a coordinate system in the lower right corner, which rotates according to the camera and minimizes the \emph{lost-in-space} problem. Besides these possibilities, advanced users can control the camera by using mouse and keyboard. We have experienced to control the camera with 3D specific devices, e.g., a 3D mouse. To see the scene simultaneously from different perspectives the user can switch to three additional lateral views by using the button in the upper left corner. Moreover, the lower left corner can be expanded to see a larger overview of the scene (see. Figure~\ref{fig:gded3d}).

These general navigation facilities make especially the orientation inside the 3D scene and furthermore the construction of generic depictions much easier. For example, the user can navigate the camera closer to an appropriate insertion context to precisely locate an insertion context. This is specifically important for the insertion of stretch intervals, which must be positioned in a way so that a container is covered by at least one interval in each dimension. To satisfy the requirement, this box must intersect the container. But to realize this requirement, the intervals must be located on a position onto the coordinate axes indicating the projection of the container onto the axes. Because of distortion effects, this position is often hard to meet. To overcome it, we realized a \emph{depth cue} that exactly shows the projection of the container on the axes by a wired box with a blue bar at the back (see see Figure \ref{fig:insCont}\subref{subfig:insCont1d} and Figure \ref{fig:gded3d}). This facilitates the insertion of the intervals, because the user can orient himself by the position of the bars. In Figure \ref{fig:gded3d} the stretch interval located on the $\mathrm{y}$-axis coincides exactly with the position of the wired box, which indicates the projection of container ``c1''. The second stretch interval on the $\mathrm{x}$-axis is smaller than the projection of ``c2'' and therefore only covers the middle of ``c2''.

The editor for generic depictions offers a way to indicate violations of requirements (containers have to be covered by at least one stretch interval in each direction and do not overlap each other) made on generic depictions. To ensure that only correct depictions are created, the requirements will be checked before generating code from the visual specification of the generic depictions. During the creation of generic depictions the user can check the program at any time by pressing a button that opens a view that lists all violations. From a language developer's point of view, requirement checks are easy to realize: For each language construct so called \emph{check functions} can be defined. Inside these functions the whole abstract structure can be accessed by using so called \emph{path expressions}.

\subsection{Code-generation}
Using the editor for generic depictions the language designer constructs three-dimensional programs to describe objects of his language. That visual specification is transformed to Java code, to be integrated in the generated language, where it is executed to draw program elements. Listing \ref{code:genDepic} shows the code generated from the generic depiction shown in Figure \ref{fig:gded3d}. For each generic depiction a class will be generated, which inherits from the \texttt{AbstractDepiction} class that contributes features, which are common for each depiction. The generated class inherits various functions to add primitives, containers, or stretch intervals. To add a container, three pieces of information are necessary: the position, the size, and the name (compare the parameters in lines 5 and 6). Stretch intervals need the declaration for which dimension they are responsible and furthermore the start and the end position (lines 8--11). The graphical primitives need the information about their position and size, a rotation value (represented by a Quaternion object) and the type of material, which together determine their visual representation. The positions of the components of the generic depictions is not extracted one-to-one from the 3D canvas but normalized to a neutral point. This neutral point is specified as the left most point for each dimension where a primitive, container, or stretch interval is located. The container inside the blue box is two pixels further inside, which can be seen on the position of the container and box in lines 5 and~13.

\begin{lstlisting}[caption={Generated code for generic depiction.},
                 label=code:genDepic, float]
class RootDepiction extends AbstractDepiction {
  public RootDepiction() {
   /* --- container --- */
   addContainer(new Vector3f(2f, 2f, 2f), new Vector3f(10f, 10f, 10f), "c1");
   addContainer(new Vector3f(23f, 3f, 3f), new Vector3f(6f, 6f, 6f), "c2");
   /* --- stretch intervals --- */
   addStretchInterval('x', 2f, 9f);
   addStretchInterval('x', 25f, 28f);
   addStretchInterval('y', 2f, 12f);
   addStretchInterval('z', 2f, 12f);
   /* --- primitives --- */
   addBox(new Vector3f(0f, 0f, 0f), new Quaternion(0f, 0f, 0f, 1f), new Vector3f(16f, 16f, 16f), Materials.blue);
   addSphere(new Vector3f(20f, 0f, 0f), new Quaternion(0f, 0f, 0f, 1f), new Vector3f(14f, 14f, 14f), Materials.green);
  }
}
\end{lstlisting}

The code generator of the generic depictions editor were be specified on the basis of the language's abstract structure. The attributes store code fragments and analysis results that are stepwise combined in order to construct the final target code. For such purpose libraries and domain-specific languages of the well-known \emph{Eli system} \cite{KPJ98} are used.

\section{Range of application}
\label{sec:rangeOfAppl}
In Figure \ref{fig:devil3dSpecification} the reader has already seen exemplary specifications for generic depictions for molecular models. Such depictions---basically representing atoms and bonds---play the role of building blocks from which molecules are be constructed. The construction of molecules is supported by a dedicated editor, which is generated by \dev. To demonstrate that our approach of specifying generic depictions is feasible for a wide range of three-dimensional languages, we present further four 3D languages and focus on the specification of generic depictions for these languages.

\begin{figure}[!ht]
  \centering
  \includegraphics[width=\textwidth]{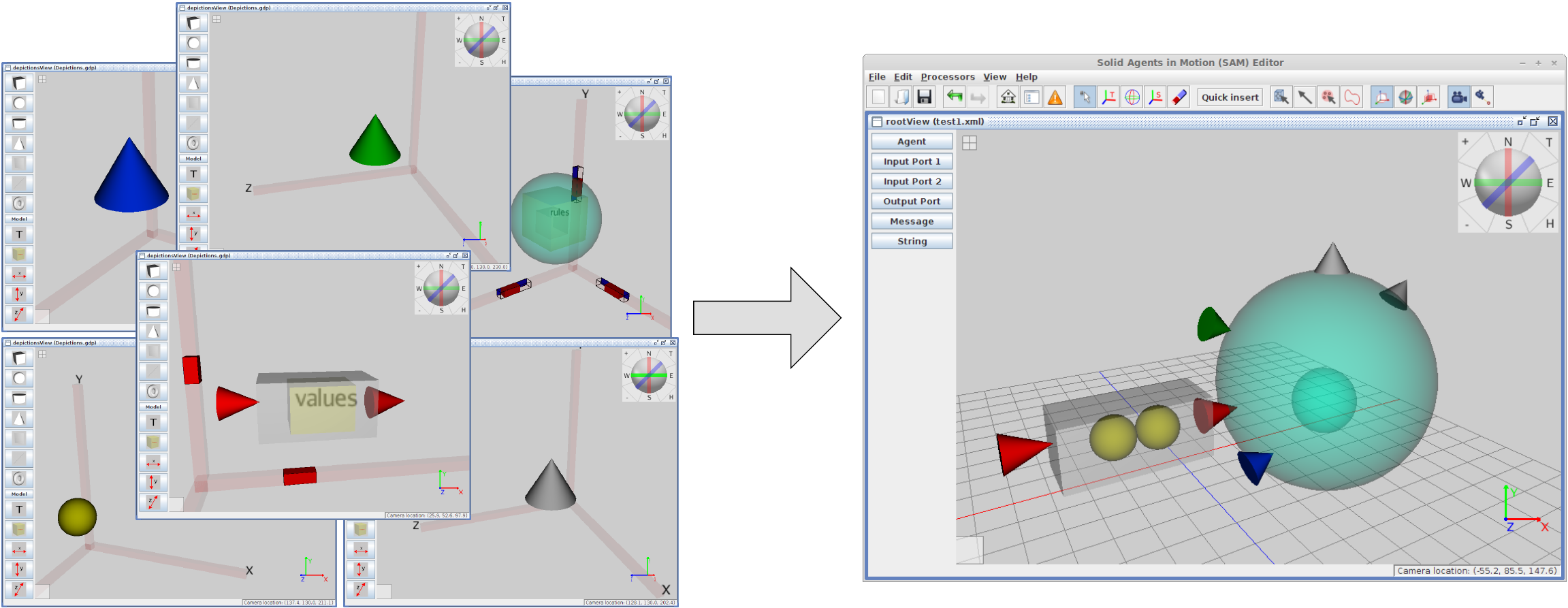}
  \caption{Generic depictions for the SAM language.}
  \label{fig:sam}
\end{figure}

Figure \ref{fig:sam} shows a specification of generic depictions for the \emph{SAM} (Solid Agents in Motion) language on the left-hand side and its application in a generated editor on the right-hand side. The SAM language \cite{GMR98} is a parallel, synchronous and state-oriented 3D language that is based on the well known 2D language \emph{Pictorical Janus}. A SAM program describes a set of agents that synchronously communicate by exchanging messages. We have specified a set of generic depictions of language constructs occurring in SAM programs. The agent is depicted by a large transparent blue sphere, which provides a container to embed rules. Furthermore, input and output ports are visualized as cones of different colors. They will be positioned in the instantiated program (right-hand side in the figure) on the surface of the agent's sphere. Messages---represented by a transparent box with two cones at its outside---are connected to the agent, too. The generic depiction for messages provides a container inside the box that can embed values of different types (for example a string value represented by a sphere) in the generated editor.

\begin{figure}[!ht]
  \centering
  \includegraphics[width=\textwidth]{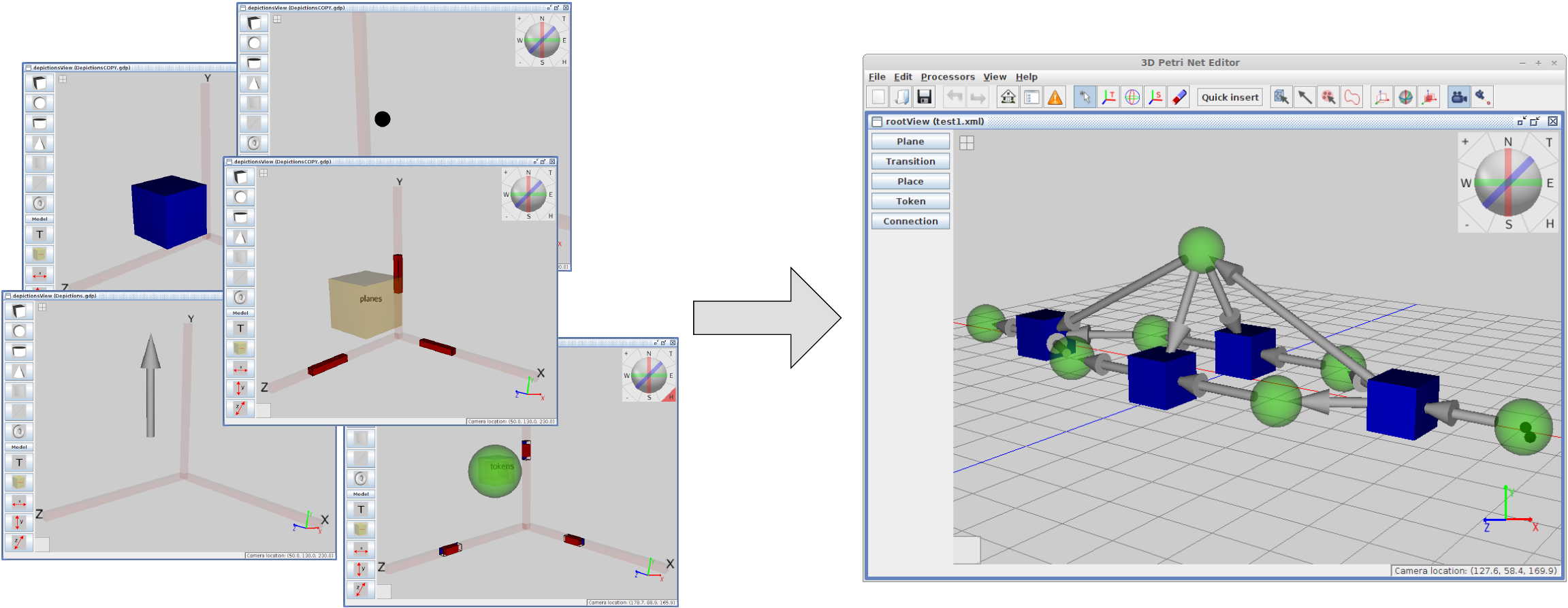}
  \caption{Visual representation of Petri Nets specified by generic depictions.}
  \label{fig:petri}
\end{figure}

As mentioned in the introduction, Petri Nets can benefit from a three-dimensional representation. R\"{o}lke \cite{Roel07} presented in his paper only the idea of 3D Petri Nets without indicating a 3D editor that allows to construct effectively in 3D. By using \dev, we have specified such an editor. To realize the visual representations for Petri Nets, five generic depictions are needed. One depiction consists of a container that embeds the whole Petri Net and adjusts its size according to the stretch algorithm when new objects are inserted. Then there are generic depictions for transitions (visualized as boxes), places, tokens (both visualized as spheres), and arrows. The depiction for places consists of a container that is located inside the sphere and can accommodate tokens. Using the generated Petri Net editor (see right-hand side in Figure \ref{fig:petri}), we have constructed a Petri Net, which has been taken from R\"{o}lke's paper. The Petri Net is partitioned in two planes: On the lower plane the control flow of the Petri Net is located and the place on the upper plane is responsible for the data-flow. The first transition writes data to this place that other transitions read. Up to now, the structure editor for 3D Petri Nets allows the creation of static Petri Nets. To simulate the execution of Petri Nets, \dev would have be extended by simulation support, as the predecessor framework DEViL \cite{CK09}.

\begin{figure}[!ht]
  \centering
  \includegraphics[width=\textwidth]{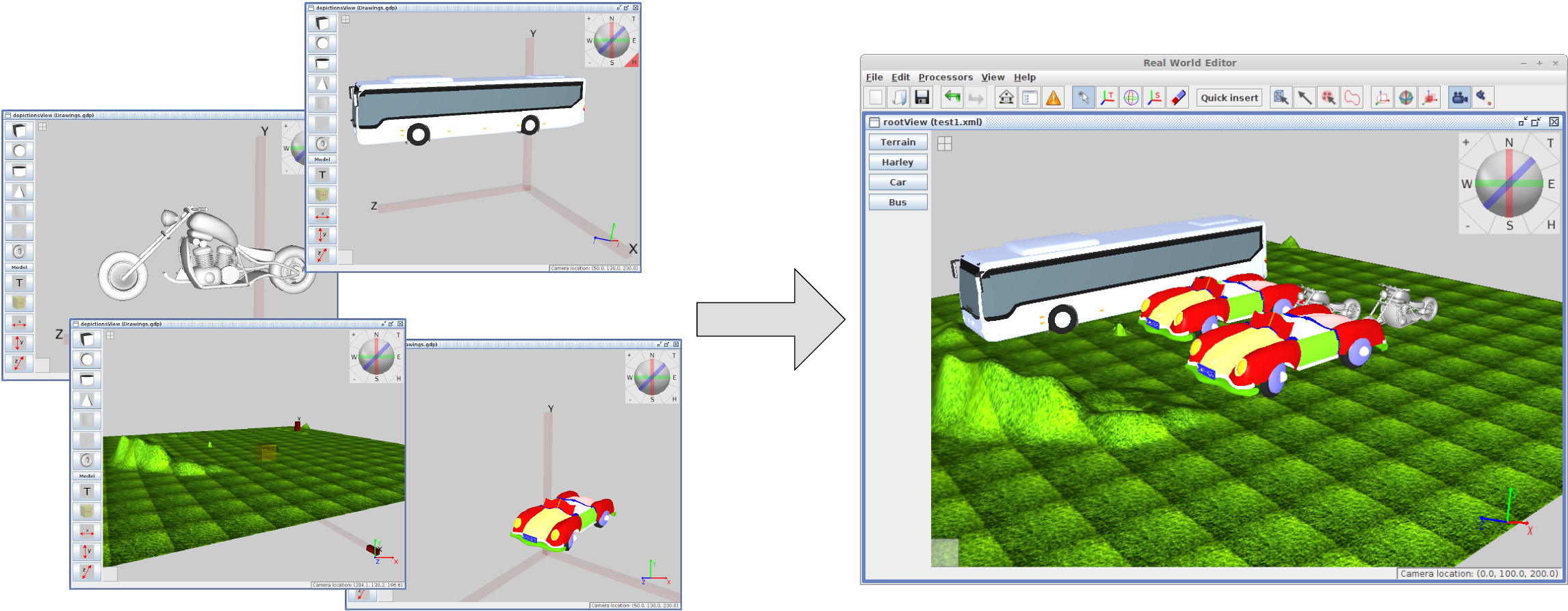}
  \caption{Generic depictions for an exemplary real-world editor.}
  \label{fig:realWorld}
\end{figure}

Figure \ref{fig:realWorld} demonstrate the specification of 3D languages, which based on real-world objects. Such objects are characterized by complex shapes, which can be modeled with tools like Blender \cite{Blender} or downloaded from libraries in the Internet containing predefined models. Such models are referenced in generic depictions. The left-hand side in Figure \ref{fig:realWorld} represents generic depictions for street vehicles as bus, car, and motor cycle. The fourth depiction shows a terrain that can be modeled with tools provided by the jMonkeyEngine \cite{jme}. In the generated editor the terrain models a base ground and the vehicles can be positioned on top of it. The positioning of the vehicles is determined by a visual pattern that arranges a set of objects on a common plane shaped surface.

\begin{figure}[!ht]
  \centering
  \includegraphics[width=\textwidth]{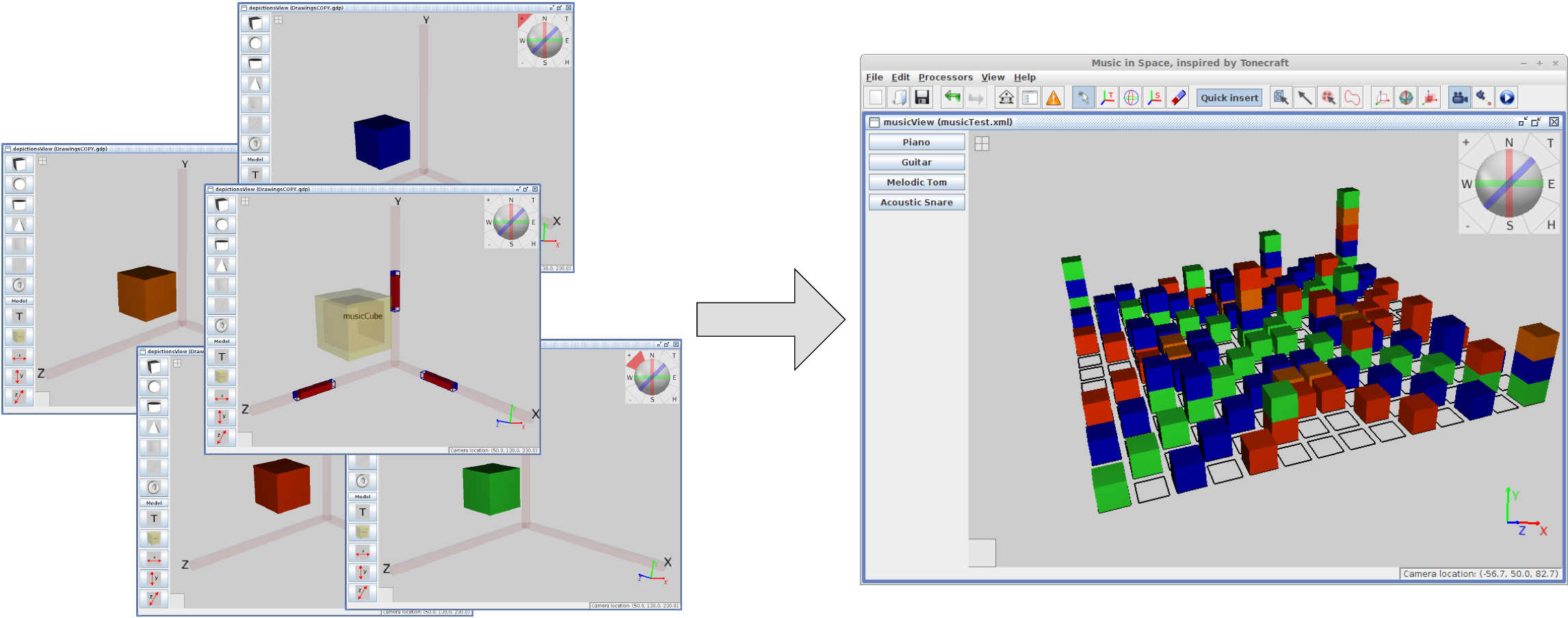}
  \caption{Music in Space editor.}
  \label{fig:music}
\end{figure}

Inspired by the web-based 3D editor ToneCraft, we have specified a \emph{Music in Space} editor with \dev. The generated editor (see Figure \ref{fig:music}) lets the user insert distinctively colored boxes representing music instruments in a matrix like area. From such a 3D program a music string according to the \emph{JFugue} music programming API \cite{jfugue} can be generated and played. The generic depictions for this editor are very simple: For each instrument a depiction representing a colored box is needed. Another depiction, which provides a container that embeds the whole program, which may be constructed according to the matrix pattern.


\section{Related Work}
\label{sec:relWork}
The basic idea of containers and stretch intervals---as used in our three-dimensional generic depictions---goes back to the VPE system \cite{Gra98} that generates 2D visual language editors. Furthermore, the idea is successfully used in the generator system DEViL \cite{SKC06}, which generates 2D language editors, too.

The idea for three-dimensional languages goes back to a publication of Glinert \cite{Gli87}. Najork \cite{Naj96} developed the first 3D language \emph{Cube}, which is semantically similar to Prolog and makes use of the data-flow paradigm. In this language all language constructs are represented by a cube. The hierarchical nesting of constructs is an inherent concept of the Cube language. In the context of generator systems for visual languages, Minas has supervised an exploration \cite{Vos09} of 3D languages in the context of his DiaGen/DiaMeta \cite{Min02, Min06} frameworks.

The work of Chung et al. \cite{CHM99} exactly addresses the topic of our paper. They have developed a tool called \emph{3DComposer} that is related to our editor for generic depictions. 3DComposer is a tool for specifying so called \emph{3Dvixels}, which can be used as building blocks for 3D applications. Such 3Dvixels can be generally used in 3D applications, which include 3D languages and also visualization tools. The usage in different applications is possible by generating reusable software components in form of JavaBeans. The construction of exemplary 3D programs is being done directly in 3DComposer by the end user. 3DComposer is not part of a generator framework, which would distinguish between language designer and language user. Hence, 3DComposer does not need concepts such as containers or stretch intervals.

The AgentCubes \cite{IRW09} system uses a mechanism to construct language object representations. For such purpose a so called \emph{Inflatable Icons Editor} is provided. It allows users to quickly draft 3D objects by drawing 2D images and turn them step by step into 3D models by using a diffusion-based inflation technique. In 3D games, which are specified with AgentCubes, these models are mostly located on the ground plane. Hence, the flat bottom side resulting from the inflation approach is not a concern.

\section{Conclusion}
\label{sec:conclusion}
In this paper we described the process of specifying generic depictions for 3D visual languages with the generator system \dev. For such purpose \dev provides the editor, which allows the language designer to specify generic depictions. This editor was also generated with \dev in a boostrapping approach. Hence, the interaction and navigation tasks, we have reported about, are available in all editors generated with \dev. For specifying generic depictions, the possibility to define containers that can embed nested constructs is particularly important. We have presented an algorithm that stretches the containers when their nested ele\-ments need more room. The language designer does not need to care about many techniques---as the stretch algorithm or the adaptation of interaction techniques to the requirements of a particular representation---because they are automatically provided by \dev.


The generic depictions editor is able to specify depictions for a wide range of 3D languages covering languages as Petri Nets or molecular models with rather simple visual representations but also languages, which consist of real-world objects that have more advanced visual representations.

\section*{Acknowledgments}
The author gratefully acknowledges the support of the German Research Foundation (Deutsche Forschungsgemeinschaft -- DFG); contract no. KA 537/6-1.
Furthermore, the author would like to thank Marius Dransfeld for the implementation of the first prototype of the generic depictions editor as well as Elena Rybka and Johann Rybka who developed interaction techniques mentioned in this paper.


\end{document}